\documentclass[conference]{IEEEtran}
\IEEEoverridecommandlockouts
\usepackage[T1]{fontenc} 
\usepackage{amsmath}
\usepackage{dsfont}
\usepackage[cmintegrals]{newtxmath}
\usepackage{bm} 
\usepackage{url}
\usepackage{mdwtab}
\usepackage{xcolor}
\usepackage[capitalise]{cleveref} 
\makeatletter
\let\MYcaption\@makecaption
\makeatother
\usepackage[font=footnotesize]{subcaption}

\makeatletter
\let\@makecaption\MYcaption
\makeatother

\usepackage[acronym,nopostdot,style=index,nonumberlist,toc]{glossaries}
\usepackage{graphicx}

\usepackage{svg}
\usepackage{mathtools}
\usepackage{makecell}

\usepackage{breqn}   

\usepackage{booktabs}
\usepackage{float}
\usepackage{tabularx}
\newacronym{mmwave}{mmWave}{millimeter wave}
\newacronym{ai}{AI}{artificial intelligence}
\newacronym{xai}{XAI}{explainable artificial intelligence}
\newacronym{embb}{eMBB}{enhanced mobile broadband}
\newacronym{urllc}{uRLLC}{ultra-reliable and low-latency communications}
\newacronym{mmtc}{mMTC}{massive machine-type communications}
\newacronym{nr}{NR}{new radio}
\newacronym{5g}{5G}{fifth generation}
\newacronym{3gpp}{3GPP}{Third Generation Partnership Project}
\newacronym{bs}{BS}{base station}
\newacronym{aoi}{AoI}{age of information}
\newacronym{jsrl}{JSRL}{jump-start reinforcement learning}
\newacronym{mac}{MAC}{multiply-accumulate operation}
\newacronym{ue}{UE}{user equipment}
\newacronym{leo}{LEO}{low-altitude earth orbit}
\newacronym{ee}{EE}{energy efficiency}
\newacronym{dci}{DCI}{downlink control information}
\newacronym{uci}{UCI}{uplink control information}
\newacronym{usc}{USC}{uplink slotted shared channel}
\newacronym{l2c}{L2C}{learning to communicate}
\newacronym{ir}{IR}{immediate reply}

\newacronym{dcm}{DCM}{downlink control message}
\newacronym{ucm}{UCM}{uplink control message}
\newacronym{pdcch}{PDCCH}{physical downlink control channel}
\newacronym{pdsch}{PDSCH}{physical downlink shared channel}
\newacronym{pucch}{PUCCH}{physical uplink control channel}
\newacronym{pusch}{PUSCH}{physical uplink shared channel}
\newacronym{rrm}{RRM}{radio resource management}
\newacronym{dlsch}{DL-SCH}{downlink shared channel}
\newacronym{bch}{BCH}{broadcast channel}
\newacronym{pch}{PCH}{paging channel}
\newacronym{ul}{UL}{uplink}
\newacronym{kpi}{KPI}{key performance indicator}
\newacronym{dl}{DL}{downlink}
\newacronym{ulsch}{UL-SCH}{uplink shared channel}
\newacronym{rach}{RACH}{random-access channel}
\newacronym{tdma}{TDMA}{time division multiple access}
\newacronym{tti}{TTI}{transmission time interval}
\newacronym{phy}{PHY}{physical layer}
\newacronym{bler}{BLER}{block error rate}
\newacronym{tb}{TB}{transport block}
\newacronym{tbs}{TBS}{transport block size}
\newacronym{per}{PER}{packet error rate}
\newacronym{prb}{PRB}{physical resource block}
\newacronym{re}{RE}{resource element}
\newacronym{rb}{RB}{resource block}
\newacronym{sr}{SR}{scheduling request}
\newacronym{sg}{SG}{scheduling grant}
\newacronym{ack}{ACK}{acknowledgement}
\newacronym{rl}{RL}{reinforcement learning}
\newacronym{ml}{ML}{machine learning}
\newacronym{mdp}{MDP}{Markov decision process}
\newacronym{ql-amc}{QL-AMC}{Q-learning based adaptive modulation and coding}
\newacronym{ql-la}{QL-LA}{Q-learning based link adaptation}
\newacronym{dqn}{DQN}{deep Q-network}
\newacronym{ma}{MA}{multi-agent}
\newacronym{marl}{MARL}{multi-agent reinforcement learning}
\newacronym{mas}{MAS}{multi-agent systems}
\newacronym{maddpg}{MADDPG}{multi-agent deep deterministic policy gradient}
\newacronym{matd3}{MATD3}{multi-agent twin delayed deep deterministic policy gradient}
\newacronym{ddpg}{DDPG}{deep deterministic policy gradient}
\newacronym{ctde}{CTDE}{centralized training and decentralized execution}
\newacronym{sdu}{SDU}{service data unit}
\newacronym{pdu}{PDU}{protocol data unit}
\newacronym{ci}{CI}{confidence interval}
\newacronym{mlp}{MLP}{multilayer perceptron}
\newacronym{relu}{ReLU}{rectified linear unit}
\newacronym{dec-pomdp}{Dec-POMDP}{decentralized partially observable Markov decision process}
\newacronym{sc}{SC}{speaker consistency}
\newacronym{ic}{IC}{instantaneous coordination}

\newglossary[nlg]{notation}{not}{ntn}{Notation}

\newglossaryentry{not:full-model}{
   name=\ensuremath{F},
   description={complete model},
   type=notation
   }

\newglossaryentry{not:decoder-model}{
   name=\ensuremath{F_{\mathrm{Dec}}},
   description={decoder model},
   type=notation
   }

\newglossaryentry{not:encoder-model}{
   name=\ensuremath{F_{\mathrm{Enc}}},
   description={encoder model},
   type=notation
   }

\newglossaryentry{not:combiner-model}{
   name=\ensuremath{F_{\mathrm{Comb}}},
   description={combiner model},
   type=notation
   }

\newglossaryentry{not:comm-system}{
   name=\ensuremath{C},
   description={communication system},
   type=notation
   }

\newglossaryentry{not:nUE}{
   name=\ensuremath{L},
   description={number of users},
   type=notation
   }

   \newglossaryentry{not:nGroups}{
   name=\ensuremath{G},
   description={number of groups},
   type=notation
   }

\newglossaryentry{not:entropy}{
      name=\ensuremath{\mathit{H}},
      description={entropy},
      type=notation
      }

\newglossaryentry{not:buffer-size}{
   name=\ensuremath{B},
   description={buffer size},
   type=notation
   }

\newglossaryentry{not:bitlength}{
   name=\ensuremath{\Upsilon},
   description={bitlength},
   type=notation
   }

\newglossaryentry{not:policy-net}{
   name=\ensuremath{\mu},
   description={policy network},
   type=notation
   }

\newglossaryentry{not:mutual-information}{
   name=\ensuremath{I},
   description={mutual information},
   type=notation
   }

\newglossaryentry{not:policy-params}{
   name=\ensuremath{\theta},
   description={policy network},
   type=notation
   }

\newglossaryentry{not:signal-overhead}{
   name=\ensuremath{\Theta},
   description={signaling overhead},
   type=notation
   }

\newglossaryentry{not:avg-packets}{
   name=\ensuremath{\lambda},
   description={average number of packets to transmit},
   type=notation
   }

\newglossaryentry{not:reward-signal}{
   name=\ensuremath{\rho},
   description={number of packets to transmit},
   type=notation
   }

\newglossaryentry{not:p-patch}{
   name=\ensuremath{p_{\mathrm{p}}},
   description={probability of having a partial observation},
   type=notation
   }

   \newglossaryentry{not:vec-query}{
   name=\ensuremath{\mu},
   description={query vector},
   type=notation
   }

   \newglossaryentry{not:generator-query}{
   name=\ensuremath{\mathcal{Q}},
   description={query generator model},
   type=notation
   }
   \newglossaryentry{not:generator-key}{
   name=\ensuremath{\mathcal{K}},
   description={key generator model},
   type=notation
   }
   
   \newglossaryentry{not:vec-key}{
   name=\ensuremath{\kappa},
   description={key vector},
   type=notation
   }

   \newglossaryentry{not:attention-weights}{
   name=\ensuremath{\mathbf{W}},
   description={attention weights},
   type=notation
   }

   \newglossaryentry{not:observation}{
   name=\ensuremath{o},
   description={intermediate observation},
   type=notation
   }

\newglossaryentry{not:episode-duration}{
   name=\ensuremath{T},
   description={maximum number of TTIs},
   type=notation
   }

\newglossaryentry{not:time-step}{
   name=\ensuremath{t},
   description={time step},
   type=notation
   }

\newglossaryentry{not:dl-vocabulary-size}{
   name=\ensuremath{D},
   description={downlink control message vocabulary size},
   type=notation
   }

\newglossaryentry{not:ul-vocabulary-size}{
   name=\ensuremath{U},
   description={uplink control message vocabulary size},
   type=notation
   }

\newglossaryentry{not:dl-message}{
   name=\ensuremath{m},
   description={downlink control message},
   type=notation
   }

\newglossaryentry{not:ul-message}{
   name=\ensuremath{n},
   description={uplink control message},
   type=notation
   }

\newglossaryentry{not:ue-action}{
   name=\ensuremath{a},
   description={UE action},
   type=notation
   }

\newglossaryentry{not:tbs}{
   name=\ensuremath{L_{\mathrm{TB}}},
   description={transport block size},
   type=notation
   }

\newglossaryentry{not:tti}{
  name=\ensuremath{T_{\mathrm{TTI}}},
  description={transmission time interval duration},
  type=notation
  }

\newglossaryentry{not:state}{
  name=\ensuremath{x},
  description={agent state},
  type=notation
  }

\newglossaryentry{not:history-len}{
  name=\ensuremath{K},
  description={length of history buffer},
  type=notation
  }

\newglossaryentry{not:episodes-train}{
  name=\ensuremath{N_\mathrm{train}},
  description={number of training episodes},
  type=notation
  }

\newglossaryentry{not:episodes-eval}{
  name=\ensuremath{N_\mathrm{eval}},
  description={number of evaluation episodes},
  type=notation
  }

\newglossaryentry{not:episodes-test}{
  name=\ensuremath{N_\mathrm{test}},
  description={number of test episodes},
  type=notation
  }

\newglossaryentry{not:repetitions}{
  name=\ensuremath{N_\mathrm{rep}},
  description={number of test episodes},
  type=notation
  }

\newglossaryentry{not:n-receptions}{
  name=\ensuremath{N_\mathrm{SDUs}},
  description={number of test episodes},
  type=notation
  }

\newglossaryentry{not:n-collisions}{
  name=\ensuremath{N_\mathrm{c}},
  description={number of collisions},
  type=notation
  }

\newcommand{\nonl}{\renewcommand{\nl}{\let\nl\oldnl}}
\let\oldnl\nl
\begin{document}

\title{Collaborative Edge Inference via Semantic Grouping under Wireless Channel Constraints}

\author{
  \IEEEauthorblockN{
    Mateus P. Mota\IEEEauthorrefmark{1},
    Mattia Merluzzi\IEEEauthorrefmark{1},
    Emilio Calvanese Strinati\IEEEauthorrefmark{1},
    }
  \IEEEauthorblockA{
    \IEEEauthorrefmark{1}CEA-Leti, Grenoble, France\\
    Email: \{mateus.pontesmota, emilio.calvanese-strinati, mattia.merluzzi\}@cea.fr
   }
   \thanks{This work was funded by the 6G-GOALS Project under the HORIZON program (no. 101139232)}
}
\maketitle

\begin{abstract}

In this paper, we study the framework of collaborative inference, or edge ensembles.
This framework enables multiple edge devices to improve classification accuracy by exchanging intermediate features rather than raw observations.
However, efficient communication strategies are essential to balance accuracy and bandwidth limitations.
Building upon 
a key-query mechanism for selective information exchange, this work extends collaborative inference by studying the impact of channel noise in feature communication, the choice of intermediate collaboration points, and the communication-accuracy trade-off across tasks.
By analyzing how different collaboration points affect performance and exploring communication pruning, we show that it is possible to optimize accuracy while minimizing resource usage.
We show that the intermediate collaboration approach is robust to channel errors and that the query transmission needs a higher degree of reliability than the data transmission itself.

\end{abstract}


\glsresetall

\glsresetall

\section{Introduction}
\label{sec:Intro}
%
Edge \gls{ai} typically involves resource-poor devices, embedded with trained \gls{ml}/\gls{ai} models, ready to output inference results on data collected from complex environments.
This differentiates from inferencing in central clouds, which benefits from huge computational resource but, at the same time, experience higher delays, increased energy consumption for data transfer, and greater data exposure.
The success of inference depends on the quality of the collected data and the model performance. When local data are corrupted by noise or missing information, collaboration with other devices through wireless communication can help improve performance.
In this direction, collaborative edge inference involves a set of devices possessing trained \gls{ml}/\gls{ai} models and performing the inference task in a cooperative way, helping each other in case of corrupted and/or missing local information.
These devices are able to communicate to share their knowledge, allowing to improve their performance, however at the cost of added communication overhead and delay.
However, the devices can perform pure local inference in case of connectivity issues, or when collaboration is not needed.
%

%
Due to its decentralized nature, collaborative inference has several advantages including but not limited to: \textit{i)} flexibility: with each device having the complete model, this paradigm allows inference even in the face of connectivity issues; \textit{ii)} low latency: by using proximity communications, latency is lower than using a centralized cloud server.  
      %
%
%
However, it comes with challenges, such as privacy, communication design and heterogeneity. 
This work explores the world of edge cooperative inference from a cross-layer perspective that entails communication, computation and application aspects, under the framework of semantic and goal-oriented communications \cite{Strinati24}, a promising paradigm towards efficiently and effectively enabling \gls{ai} services in 6G.\\
\textit{\textbf{Related works}} \,
A central challenge in collaborative inference is determining which features to share and how to select collaboration partners.
The problem studied in this work is similar to collaborative perception \cite{han2023collaborative}, however, we consider the model as fixed instead of learned.
In \cite{liu2020when2com}, a collaborative perception framework that dynamically decides when and with whom to communicate based on a query-key handshake to generate a learned communication graph. 
A similar framework is used in \cite{huang2023semdas} for semantic data sourcing, where an edge server broadcasts a semantic query to request relevant data from its sources for a given task.
This is further extended to random access in \cite{kalor2023random}, where the matching between the edge server semantic query and a device key is used to determine the transmission probability.
None of these papers analyze the effect of channel noise during communication, e.g., pure noise or bit/packet-level errors/erasures.
\\
%
\textit{\textbf{Contribution}} \,
In this paper, we assume the task is performed by each edge device, i.e., closer to the framework of \cite{liu2020when2com}, but analyzing it under wireless system constraints such as errors introduced by wireless communication. We study the effect of channel errors in the semantic communication graph grouping \cite{liu2020when2com} and the communication design choices of this method.
As such, the goals of this work involve:
\begin{itemize}
      \item Studying the effect of the channel in the collaborative inference problem.
      \item Studying the best splitting point and the trade-offs involved in its choice.
      \item Analyzing the accuracy-communication cost trade-off.
\end{itemize}

%
%



This work is structured as follows.
\Cref{sec:system-model} describes the system model, \Cref{sec:solution} introduces the framework for communication grouping based on semantic matching.
Finally, \Cref{sec:results} details the experiments performed and discusses their results.
The work is then concluded in \Cref{sec:conclusion}.


\section{System Model}
\label{sec:system-model}

\begin{figure}[t]
      \centering

      \includegraphics[width=0.96\columnwidth]{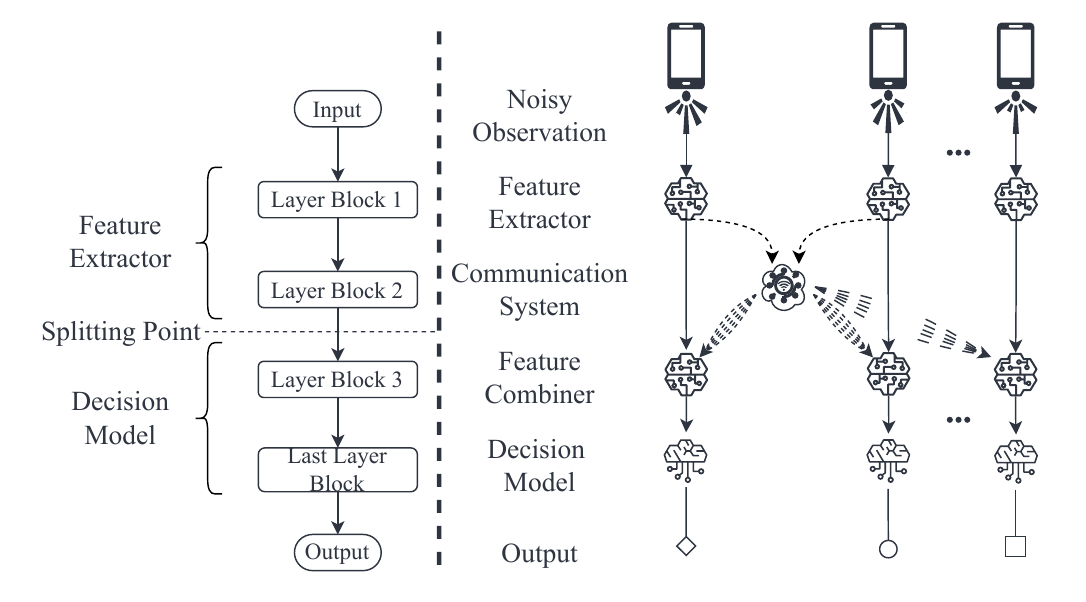}
      \caption{
            System model scheme for the proposed collaborative inference problem.
      }
      \label{fig:system-model}
\end{figure}

We consider a system composed of \gls{not:nUE} devices empowered with \gls{ai} capabilities, in this case inference models.
Each device $i$ performs inference using a pre-trained model $\gls{not:full-model}$ on input $x_i$, e.g., an image collected through a camera or other modality data.
Without loss of generality, this model can be split into two parts, a feature extractor $\gls{not:encoder-model}$ and a decision model $\gls{not:decoder-model}$, as in \cref{fig:system-model}, such that 
\begin{dmath}
      \gls{not:full-model} (x) = \left( \gls{not:decoder-model} \circ \gls{not:encoder-model} \right) \, (x)
\end{dmath}.

We assume that the \gls{not:nUE} devices can be clustered in $\gls{not:nGroups}$ groups based on the input data they generate at a given time instant, with the devices in the same group generating the same data. 
As such, if devices $i$ and $j$ are in group $g$, $x_i = x_j = x_{g}$.
Without loss of generality, it is assumed that each group is composed of a number of devices and that the groups are randomly constructed, as such the devices are not aware of the identity of other devices in their group.
The ultimate goal is to mutually discover these identities to improve local inference performance, with low overhead for the network, i.e. without the need of directly sharing data.

With probability $\gls{not:p-patch}$, a device has access only to partial or noisy observation of the true data, i.e., the true group observation, $x_g$,  given by $\hat{x}_i = M^i(x_g)$, with $i \in \left[ 1 \, .. \, \gls{not:nUE} \right]$ and $g \in \left[ 1 \, .. \, \gls{not:nGroups} \right]$. 
Otherwise, with probability $1-\gls{not:p-patch}$, the device has access to the true observation, $x_g$.
All devices are able to communicate by sharing the output of the first part of the model $\gls{not:encoder-model} \, (\hat{x}_i)$, which is then shared with other devices (at least one) over a wireless communication link $\gls{not:comm-system}$. We intentionally keep the notion of communication system general. The latter can be translated into a wireless link, end-to-end link to edge, or whatever transformation and transportation performed on data. As an example, in this work, $\gls{not:comm-system}$ is represented by a wireless packet erasure channel.
This results in a possibly corrupted version of data at the receiver.

For a device $i$, the information received by other devices in its group is instrumental to improve local inference performance, especially in the case of noisy or corrupted local observation.
However, this requires the discovery of the devices belonging to the same group, as well as good enough channel conditions.
%
%
Whenever a generic message $o_j$ is transmitted by device $j$, the message $y_c^{ij}$ received by device $i$, after passing through said communication system is denoted by 
\begin{dmath}\label{eq:rx_message}
    y_c^{ij} = \gls{not:comm-system} \left\{  \gls{not:observation}_i \right\}
\end{dmath},
where we use the notation $\{\cdot\}$ to denote a system, rather than a function.


The devices need to aggregate the information received by their created group.
This is performed by the feature combiner, $\gls{not:combiner-model}^i$ at device $i$.
Denoting by $\mathbf{y}_c^i$ the aggregated information received by device $i$ from the other devices in the group, the output of the feature combiner is
\begin{dmath}\label{eq:feat_comb}
    y^i_g = \gls{not:combiner-model} \left( \gls{not:encoder-model} ( \hat{x}_i ) , \mathbf{y}_c^i \right)
\end{dmath}.

Finally, the combined information is fed to the decision model to provide the inference result at device $i$:
\begin{dmath}
    y_d^i = \gls{not:decoder-model} (y^i_g)
\end{dmath}.
This whole procedure is illustrated in \cref{fig:system-model}.

The end goal of this setting is to allow collaborative inference, improving the overall accuracy by means of sharing intermediate information over wireless.
For this reason, the devices are allowed to use the communication system, not only to share their feature information, but also to identify potential devices with the information relevant as per the scope of improving their inference task performance.
This needs weighting their contribution in the feature combiner.
Final performance of this collaboration depends on: (i) the effectiveness of group creation, (ii) the wireless channel quality between collaborating devices, and (iii) that the splitting point of model $\gls{not:full-model}$.
The latter not only affects performance in terms of accuracy, but also the communication system, since different splitting points provide different feature size.

\subsection{Communication System}

Sidelink communication between devices can happen in three different ways:
\begin{itemize}
      \item \textbf{Unicast:} One-to-one communication. Data is transmitted to a single device.
      \item \textbf{Multicast:} One-to-many communication. Data is transmitted to a dedicated set of devices in the area.
      \item \textbf{Broadcast:} One-to-all communication. Data is transmitted to all devices in the broadcast area.
\end{itemize}
Unicast and multicast communication need network connection, since they use the uplink to request communication, i.e. request to join a group.
The solution studied in this paper relies on both multicast for the transmission of \textit{semantic queries} and unicast to exchange intermediate observation.

%
The communication system $C$ is a sidelink packet erasure channel, where a \gls{tb} is incorrectly received with a probability defined as \gls{per}.
The transmitted data is divided into \glspl{tb} of size $N$, which for simplicity we assume to be the number of floating-points values.
The system is assumed to have constant resources, such that $N$ and \gls{per} are fixed.
Also, we do not assume retransmissions in this work, so that if a \gls{tb} is not received then its values are filled with a default value, e.g. $0$.

\subsection{Key Performance Indicators}
%
%
%
Semantic and goal-oriented communication goes beyond classical wireless communication metrics towards application success.
Focusing on image classification, we consider accuracy as key performance indicator at the application level.
%
%
%
%
%
\begin{table*}[t]
      \begin{minipage}[t]{\columnwidth}
            \caption{Simulation Parameters}
            \centering
            \label{tab:sim-params}
            \begin{tabularx}{0.95\columnwidth}{l c c}
            \toprule
            \textbf{Parameter}     & \textbf{Symbol}  & \textbf{Value}           \\
            \midrule
            Number of devices      & \gls{not:nUE}         &  $16$                    \\
            Number of groups       & \gls{not:nGroups}     &  $4$                     \\
            Packet error rate      & \gls{per}             &  $10^{-1}$               \\
            Transport block size  (\# of floating-points)  & \gls{not:tbs}  &  $40$   \\
            Filling value for packet losses       &        &  $0.0$                   \\
            Query size             & Q                     &  $64$                    \\
            Key size               & K                     &  $1024$                  \\
            Patch scale            &                       &  $0.4$                   \\
            Probability of partial observation  &   \gls{not:p-patch} &  $0.8$        \\
            \bottomrule
            \end{tabularx}
      \end{minipage}\hfill 
      \begin{minipage}[t]{\columnwidth}
            \caption{Training Parameters}
            \centering
            \label{tab:train-params}
            \begin{tabularx}{0.9\columnwidth}{l c}
            \toprule
            \textbf{Parameter}       & \textbf{Value} 	            \\
            \midrule
            Batch size          & $64$                             \\
            Number hidden layers      & $ 2 $      \\
            \# of neurons per hidden layer in \gls{not:generator-key} and \gls{not:generator-query}     & $ \left[ 256, 128 \right] $      \\
            Activation function of hidden neurons    & ReLU      \\
            Optimizer algorithm     & Rectified Adam    \\
            \# of epochs         & $60$                     \\
            Learning rate        & $ 10^{-5}$                     \\
      
          \bottomrule
          \end{tabularx}
        \end{minipage}
      \end{table*}

%
%
However, application performance are typically to be traded off with cost, e.g., in terms of communication and computation.
Here, focusing on the sidelink, we consider the average resource usage as KPI.
%
The latter is computed as the average resource usage of the system, namely the number of sidelink transmissions resulting from the optimized device grouping (i.e., the communication graph).
%

\section{Semantic matching-based grouping}
\label{sec:solution}


Given the above task, this work leverages an attention-based mechanism similar to \cite{liu2020when2com} to identify: i) whether a device needs extra information for inference due to corrupted local data, and ii) the set of devices to collaborate with, in case of bad quality local data.
%
%
Adapting the framework to our system, device $i$ compresses its observation, obtaining an intermediate representation $\gls{not:observation}_i = \gls{not:encoder-model} \, (\hat{x}_i)$.
Then, it generates: \textit{i)} a low-dimensional query vector $\gls{not:vec-query}_i$ and \textit{ii)} a key vector $\gls{not:vec-key}_i$:
\begin{gather}
      \gls{not:vec-query}_i = \gls{not:generator-query} (o_i ; \theta_q ) \, \text{,} 
      \\
      \gls{not:vec-key}_i = \gls{not:generator-key} (o_i ; \theta_k) \, \text{,}
\end{gather}  
\noindent where $\gls{not:generator-key}$ and $\gls{not:generator-query}$ are two neural networks parametrized by $\theta_k$ and $\theta_q$, respectively.
The query is transmitted to all other devices, while the key is kept local.
The query received by device $j$ from device $i$ is $\hat{ \gls{not:vec-query}}^j_i=C\{\mu_i\}$.

All devices receive the queries of all others (multicast), and uses their keys to compute a matching score through scaled general attention \cite{luong2015effective}. Then, it exchanges data (unicast).
We denote by $m_{i j}$ the matching score for device $j$ receiving query from device $i$, which reads as 
\begin{equation}
\label{eq:general-attention}
      m_{ij} = \frac{{\gls{not:vec-key}_i}^{\intercal} \gls{not:attention-weights} \hat{\gls{not:vec-query}}^j_i}{\sqrt{K}} \, \text{,}
\end{equation}
\noindent where $\gls{not:attention-weights} \in \mathbb{R}^{Q \times K}$ is a learnable parameter to match the query size, $Q$, and the key size, $K$.

%
All the matching scores are used to construct a matching matrix $\mathbf{M}$ by using a row-wise softmax, with elements $\bar{m}_{ij}$.
The latter is used to construct the communications graph, as its values $\bar{m}_{i j}$ represent how relevant the information of device $i$ is for device $j$.
Once the groups are created, the devices share the actual data (or, their intermediate representation), which might have high dimensionality compared to the queries.
To avoid high communication overhead, $\mathbf{M}$ can be pruned with threshold $\rho$, i.e., $\bar{m}_{ij}^\rho=\bar{m}_{ij}\cdot\mathbf{1}\{\bar{m}_{ij}\geq\rho\}$, where $\mathbf{1}\{\cdot\}$ denotes the indicator function.
$\mathbf{M}$ is also used to combine features (cf. \eqref{eq:feat_comb}) according to the following weighted average:
\begin{equation}
      y^i_g = \sum_{j=1}^{\gls{not:nUE}} \bar{m}_{i j}^\rho y_c^{ij} \, \text{.}
\end{equation}
where $y_c^{ij}$ is the received version of the intermediate data, as per the definition in \eqref{eq:rx_message}.

In this work, we consider image classification as application. As such, training is performed by computing the cross-entropy loss between the true label and the predicted label, $y_d^i = \gls{not:decoder-model} (y^i_g)$.
It is important to highlight that, differently from \cite{liu2020when2com}, only the query generator $\gls{not:generator-query}$, the key generator $\gls{not:generator-key}$ and the attention weights $\gls{not:attention-weights}$ are learned.
As such, the encoder model $\gls{not:encoder-model}$ and the decoder model $\gls{not:decoder-model}$ are assumed to be pre-trained and their parameters frozen, while only the modules needed for the communication need to be trained.
As a consequence, the learned encoder and decoder models are shared across all devices.
Note that decentralized training is also possible, with the result of different models for each device. However, this increases the computational cost.
Differently from \cite{kalor2023random}, training to learn key and query takes into consideration the channel effects, i.e.  packet losses, instead of only considering it during execution.
This introduces robustness to noise and thus the possibility of improving inference performance through wireless collaboration, without the need for high communication reliability, e.g., through retransmissions.
\section{Experiments and Discussion}
\label{sec:results}

\subsection{Simulation details and parameters}

Image classification is performed on the Imagenette dataset \cite{Howard_Imagenette_2019}.
The pre-trained model is the MobileNetV3-Small \cite{howard2019searching}, initialized with its default weights from training in the ImageNet dataset and then fine-tuned to the Imagenette dataset.
The partial observability is modeled by applying a white patch in a random position of the image, with the ratio between the white patch size and the image size being $0.4$. In other words, $40$\% of the image is locally missing at the device, if the latter belongs to the set of devices with corrupted data.
%
%
We do not consider generalization in the following results, as the system is trained for each combination of parameters considered.

Unless otherwise stated, it is assumed that the query vector is transmitted reliably, in an error-free manner and given its size compared with the intermediate feature vector, its contribution to the resource usage is not neglected. 
The intermediate feature vector is thus the only information affected by channel errors.
%

We highlight that computation overhead added by the key-query and attention modules is negligible with respect to the base model (encoder and decoder).
With the architectures used, the total number of flops for running MobileNetV3-Small is $55$M \glspl{mac}, while the key and query networks amount to only $1.25$M and $1.37$M \glspl{mac}, respectively, and the attention module to $0.07$M \glspl{mac}.


The semantic grouping solution is compared with two other benchmark solutions:
\begin{itemize}
      \item \textbf{Local inference:} only the local observation is used for inference, possibly on corrupted data. This represents the non-collaborative case.
      \item \textbf{Noiseless:} collaboration is performed without wireless channel impairments. This represents a performance upper bound
      \item \textbf{Naive grouping:} the observation of all devices are averaged with the same weights, not based on the semantic data relevance.
\end{itemize}
%
We analyze numerical results by varying: i) the packet error rate, ii) the DNN splitting point for extracting data for collaboration, iii) pruning threshold, and ultimately the channel errors on the queries.
\begin{figure}[!tb]
      \centering
            \includegraphics[width=0.9\columnwidth]{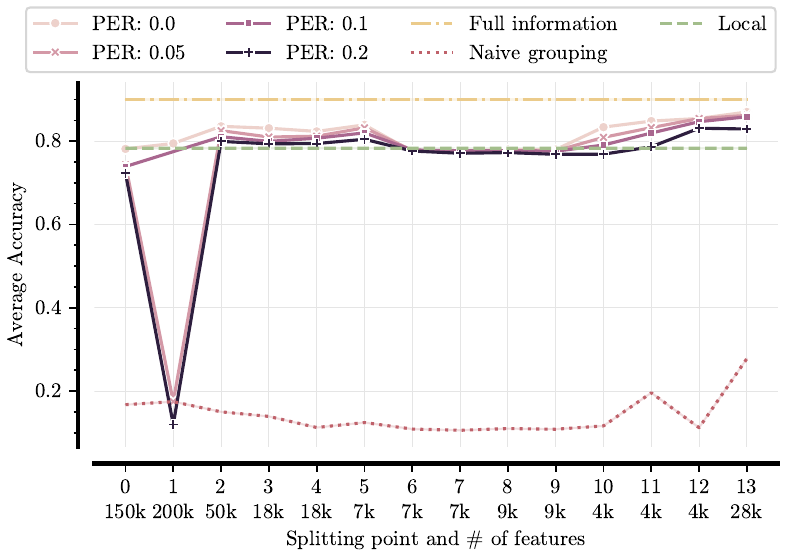}
            \caption{Accuracy and \# of features when varying the splitting point.
            }
            \label{fig:splitting-result}
      \end{figure}

\begin{figure}[!tb]
      \centering
            \includegraphics[width=0.9\columnwidth]{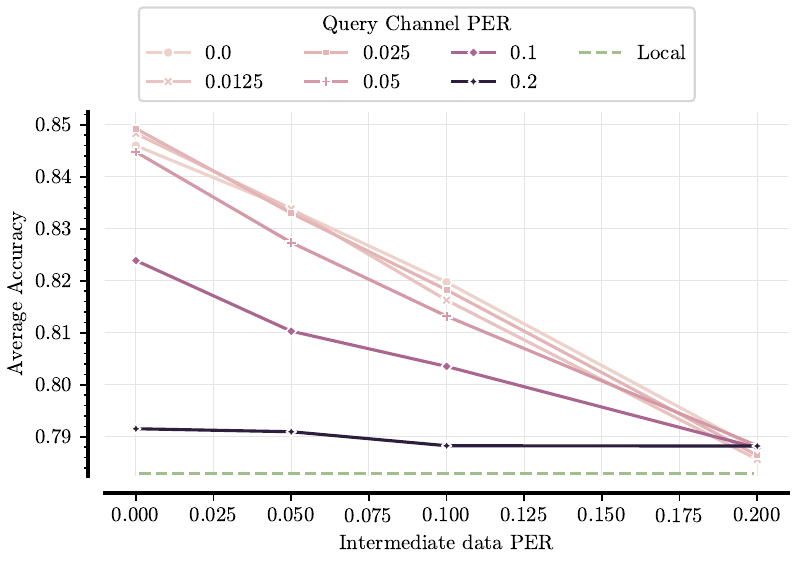}
            \caption{Accuracy for different query and intermediate data \glspl{per}. 
            itting point: 11.            }
            \label{fig:per-result}
  \end{figure}

\subsubsection{Splitting point choice}
%
First, we analyze the effect of the splitting point choice, comparing the semantic grouping solution trained with different \glspl{per} with the baselines in \cref{fig:splitting-result}.
Namely, we show the accuracy as a function of the splitting point, for different \glspl{per}. The size of the intermediate feature observation is highlighted below each splitting point in the abscissa.
%

From \cref{fig:splitting-result}, we can draw the following conclusions:
\begin{itemize}
      \item Robustness to noise depends splitting point, as initially observed in \cite{Binucci24}: Some splitting points (e.g., $4$ and $12$) show a strong robustness, where there is little performance variation, while other points are more susceptible to channel errors, such as points $0$ and $1$;
     \item Naively grouping devices leads to extremely poor performance, as it does not take into account the semantic relevance of data when collaborating;
      \item It is possible to improve accuracy through collaboration while transmitting less information, as shown by the increased performance in terms of accuracy when going deeper before collaborating. This helps reducing the resource usage per transmission.
\end{itemize}
The above conclusions imply that the \gls{phy} can relax its reliability levels as long as an appropriate semantic extraction can be performed.
We also note that naive grouping does not perform well since it fails to filter the data based on its relevance to a device.

\subsubsection{Channel effect on query and data transmission}

We conclude that \gls{phy} reliability is not necessary when semantic information is exchanged. However, in the previous result, we considered error-free communication for the query. 
%
%
In \cref{fig:per-result}, we plot the accuracy as a function of the \glspl{per} during intermediate data transmission, for different \glspl{per} during query transmission.
The system is trained for each combination of query and data channel \glspl{per}.

We can notice the severe effect of query errors, as the accuracy drops nearly twice as much with an increase in the query channel \gls{per} compared to a similar increase in the intermediate data channel \gls{per}.
This suggests that the query needs a reliable transmission scheme, compared to the data (or intermediate representation) itself.
We can conclude that semantic representation helps communication robustness, but device grouping needs reliable query exchange to achieve acceptable performance.
However, it should be noted that, given the small size of the query, increased communication reliability effort does not impact system cost as data transmission itself.
\begin{figure}[!tb]
      \centering
      \begin{subfigure}[t]{0.94\columnwidth}
            \centering
            \includegraphics[width=\textwidth]{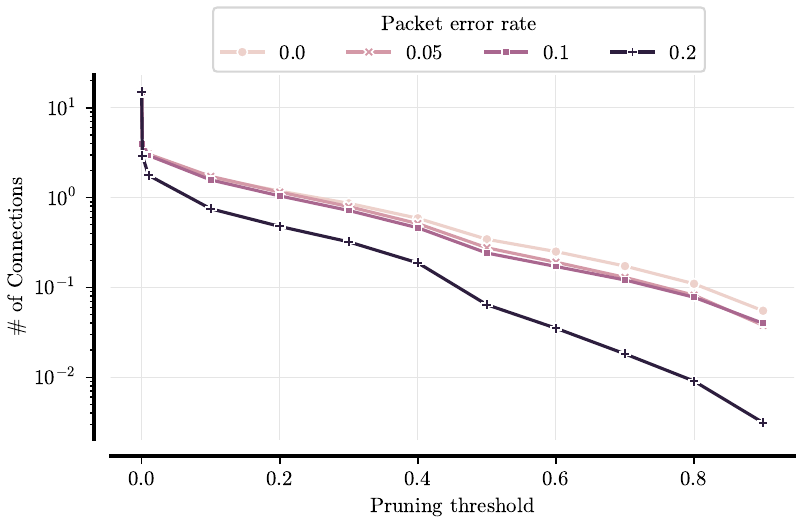}
            \caption{Effect of the pruning threshold on the average number of sidelink connections per device.}
            \label{fig:conn-thr}
            \end{subfigure}%
            \vspace{8pt}
            \begin{subfigure}[t]{0.94\columnwidth}
                  \centering
                  \includegraphics[width=\textwidth]{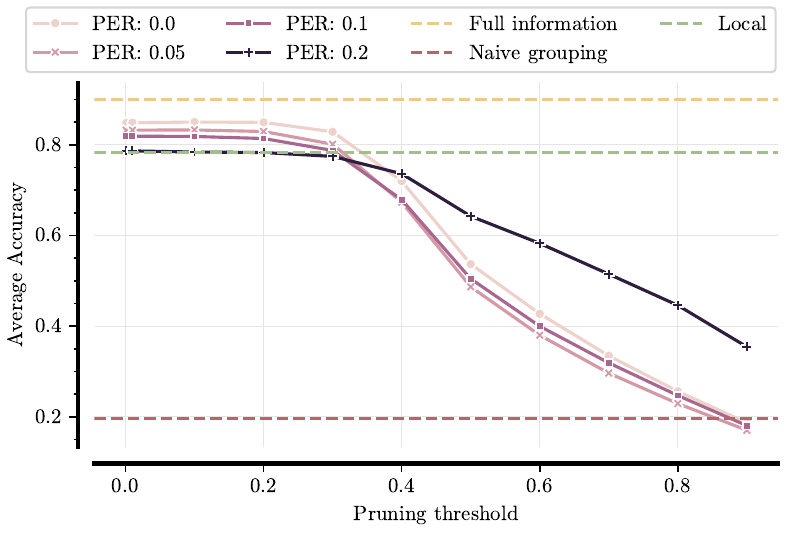}
                  \caption{Effect of the pruning threshold on the average accuracy.}
                  \label{fig:acc-thr}
            \end{subfigure}
            \caption{Studying the effect of the communication pruning. The lowest thresholds are $\left[ 0, 0.001, 0.01 \right]$. Splitting Point: 11}
            \label{fig:results-thr}
      \end{figure}

\subsubsection{Communication pruning effect}
%
We now analyze the effect of the communication pruning, which reduces the communication by only transmitting information if the matching score is above a certain threshold.
These results are shown in \cref{fig:results-thr}.
In \cref{fig:conn-thr}, we plot the average number of sidelink connections per inference task, as a function of the pruning threshold $\rho$.
Whereas, \cref{fig:acc-thr} shows the corresponding accuracy, again as a function of $\rho$.
\cref{fig:conn-thr} shows that the lowest non-negative threshold already provides significant communication reduction.
This is thanks to the fact that some of the elements of the matching matrix are very close or equal to zero. Naturally, this does not reduce accuracy, as shown in \cref{fig:acc-thr}, because only lowly weighted collaborations are pruned.
\cref{fig:conn-thr} also shows that bigger pruning reduces communication even more when the channel conditions worsen.
This implies that the matching scores have lower variation under strong channel impairments, which insinuates that the system learns to rely on information from more sources when \gls{per} is high than when it is low, in which case the system instead relies on few relevant sources.
Comparing \cref{fig:conn-thr} and \cref{fig:acc-thr}, we can conclude that it is possible to reduce communication effectively without affecting accuracy.
However, if communication is heavily reduced, the degradation in performance can even overcome local performance with corrupted data.
\section{Conclusions and Perspectives}
\label{sec:conclusion}

We investigated the impact of wireless communication impairments on collaborative inference in edge AI systems, focusing on semantic grouping, communication efficiency, and model partitioning under varying channel conditions.
By leveraging a key-query mechanism for selective feature exchange, we demonstrated that adaptive communication strategies can significantly improve inference accuracy while minimizing resource usage. 
We showed that the semantic grouping solution is robust to channel errors. Nevertheless, the query channel requires more reliability than the data channel.  
Furthermore, it is possible to improve the accuracy of the inference task while reducing the communication cost by appropriate choice of the splitting point and communication graph pruning.

Our findings provide practical guidelines for designing scalable and communication-aware edge AI deployments and operations.
Future research directions include extending the framework to consider the channel information in the matching matrix, so that the model can handle different channel conditions.
Another interesting perspective is a query-aware data transmission, such that the query is used to extract the more relevant features to be transmitted.
In terms of evaluation, a comparison on different tasks such as data sourcing \cite{huang2023semdas} and semantic segmentation \cite{liu2020when2com} can provide better insights into the wireless effects across different tasks. 


      
\bibliographystyle{IEEEtran}
\bibliography{IEEEabrv, ref.bib}

\end{document}